\documentclass[prl,twocolumn,showpacs,preprintnumbers,superscriptaddress,amsmath,amssymb]{revtex4-1}

\usepackage{graphicx}
\usepackage{array,multirow,amsmath,amsfonts}
\usepackage{color}

\def\EF{$E_\textrm{F}$}
\def\kF{$k_\textrm{F}$}

\def\4kBT{$4k_\textrm{B}T$}
\def\B6{B$_6$}
\def\etal{$et\ al.$}
\def\deg{$^\circ$}
\def\Gbar{$\overline{\Gamma}$}
\def\Xbar{$\overline{\textrm{X}}$}
\def\Ybar{$\overline{\textrm{Y}}$}

\def\Mbar{$\overline{\textrm{M}}$}
\def\110b{$<$110$>$}
\def\invA{\AA$^{-1}$}

\definecolor{dkgreen}{rgb}{0.21,0.49,0.16}

\begin{document}
\preprint{}

\title{Consistency of Photoemission and Quantum Oscillations for Surface States of SmB$_6$ }

\author {J. D. Denlinger}
\email[]{Email address: jddenlinger@lbl.gov}
\affiliation{Advanced Light Source, Lawrence Berkeley Laboratory, 
	Berkeley, CA 94720, USA}
\author {Sooyoung Jang}
\affiliation{Advanced Light Source, Lawrence Berkeley Laboratory, 
	Berkeley, CA 94720, USA}
\affiliation{Department of Physics, Pohang University of Science and Technology, 
	Pohang 37673, Korea}
\author {G. Li}  
\author {L. Chen}  
\author {B. J. Lawson}  
\author {T. Asaba}   
\author {C. Tinsman}  
\author {F. Yu}   
\author {Kai Sun} 
\author {J. W. Allen}  
\author {C. Kurdak}   
\affiliation{Department of Physics, Randall Laboratory, 
	University of Michigan, Ann Arbor, MI 48109, USA}  
\author {Dae-Jeong Kim} 
\author {Z. Fisk}   
\affiliation{Department of Physics and Astronomy, 
	University of California at Irvine, Irvine, CA 92697, USA}
\author {Lu Li}
\email[]{Email address: luli@umich.edu}
\affiliation{Department of Physics, Randall Laboratory, 
	University of Michigan, Ann Arbor, MI 48109, USA} 
	
\date{\today}

\begin{abstract}

The mixed valent compound Sm\B6\ is of high current interest as the first candidate example of topologically protected surface states in a strongly correlated insulator and also as a possible host for an exotic bulk many-body state that would manifest properties of both an insulator and a metal. Two different de Haas van Alphen (dHvA) experiments have each supported one of these possibilities, while angle resolved photoemission spectroscopy (ARPES) for the (001) surface has supported the first, but without quantitative agreement to the dHvA results.  We present new ARPES data for the (110) surface and a new analysis of all published dHvA data and thereby bring ARPES and dHvA into substantial consistency around the basic narrative of two dimensional surface states. 

\end{abstract}



\maketitle


Samarium hexaboride (Sm\B6) is well established as a bulk mixed valent insulator 
at low temperature \cite{Riseborough00}.  It is of high current interest as the first candidate example of a new class of strongly correlated topological insulators \cite{Dzero10,Takimoto11,Lu13,Wolgast13,Kim13}, which implies topologically protected metallic surface states in a bulk insulating gap.  There is now consensus that angle resolved photoemission spectroscopy (ARPES) observes a bulk gap and in-gap (001) surface states \cite{Xu13,Neupane13,Jiang13,Min14,Denlinger13b,Hlawenka15} although there is debate  \cite{Hlawenka15} as to their topological origin or not.  Quantum oscillations in magnetization (de Haas-van Alphen (dHvA) effect) by Li \etal~\cite{Li14} have also supported the existence of metallic two dimensional (2D) states on (001) and (011) surfaces, but with essentially no detailed agreement to ARPES.  In particular, the most prominent and well established ARPES surface state has not been reported in any published dHvA study.

Sm\B6\ is also of high current interest for the exciting possibility of a yet more exotic bulk electronic structure \cite{Tan15,Knolle15,Baskaran15,Zhang15}  that could manifest 3D bulk quantum oscillations normally associated with a metallic Fermi surface in spite of being a bulk insulator.  At the moment then, the two powerful experimental electronic structure techniques of dHvA and ARPES have failed to deliver a consistent picture as to the most basic aspects of the electronic structure of Sm\B6.  In this paper we present a new analysis of all published dHvA data and new ARPES data for the (011) surface and thereby bring ARPES and dHvA into substantial consistency around the basic narrative of 2D surface states.  Not only is there no necessity to invoke a 3D model for the dHvA data, but such a model is actually inconsistent with the dHvA data of  Li \etal.


In the first dHvA study by Li \etal\ experimental orbit frequencies up to 1 kT were reported for angular dependence in the cubic $x$-$z$ plane over a wide 210\deg\ range encompassing (001),(101),(100),(10$\overline1$) and (00$\overline1$)) directions of the magnetic field,  and were interpreted with a 2D model \cite{Li14}.  In the second dHvA study by Tan \etal\   \cite{Tan15}, orbit frequencies were reported up to 15 kT with angular dependence in a diagonal plane over a 90\deg\ range from (001)-(111)-(110), and were interpreted with a bulk Fermi surface model very similar in size and shape to what is very well known for cubic La\B6 \cite{Ishizawa77,Harima88,Onuki89ce}.  There, the dominant feature is six Brillouin zone (BZ) X-point ellipsoids that are large enough to overlap because the trivalency of La yields one La $d$-band conduction electron per unit cell after entirely filling B $p$-states.  The overlap generates neck-related FS orbit frequencies \cite{Ishizawa77} as well as much smaller orbits arising from intra-neck ellipsoids along the \110b\ axes \cite{Harima88}. 

A general caveat with the La\B6\ model is that La\B6\ has an odd number of electrons per cell whereas the number in Sm\B6\ is even. An early hint of difficulty is that the dominant amplitude feature in both Sm\B6\ dHvA measurements is the oscillation pattern in the 0.3-1 kT frequency range \cite{suppl}, whereas the dominant amplitude features in La\B6\  \cite{Goodrich09}, arising from the large X-point ellipsoids, occur for 8-10 kT.  Other problems related to the much reduced $d$-band filling that results from the reduced (mixed) valence of Sm are pointed out below. Finally, as discussed in detail in the Supplement \cite{suppl}, there are serious conceptual problems with the additional assumption by Tan \etal\ of universal trivalent Fermi surface features between the relatively simple FS topology of La\B6\ and the complex zone-folded FS topology of antiferromagnetic Pr\B6\ \cite{Onuki89ce,Kasuya96}. Thus their transference of orbit labeling from La\B6\ to Pr\B6\ and then to Sm\B6\ is very problematic.

We first present a comparison of the dominant amplitude 0.3-1 kT dHvA angular dependences in the two studies and answer the question as to whether a 2D or a 3D model fits better.  The first dHvA study also reported supplementary diagonal plane angular dependence up to 1 kT which can be directly compared to the second study and to the proposed 2D and 3D FS orbit models. Such a comparison has been made \cite{Erten15} with the conclusion that while the two data sets are in agreement with each other, it is not possible to distinguish if a 2D or a 3D model fits better.  A much sharper conclusion can be drawn by including in the analysis the detailed $x$-$z$ plane angular dependent data set from the first dHvA study.

\begin{figure}[t]
\begin{center}
\includegraphics[width=8.5cm]{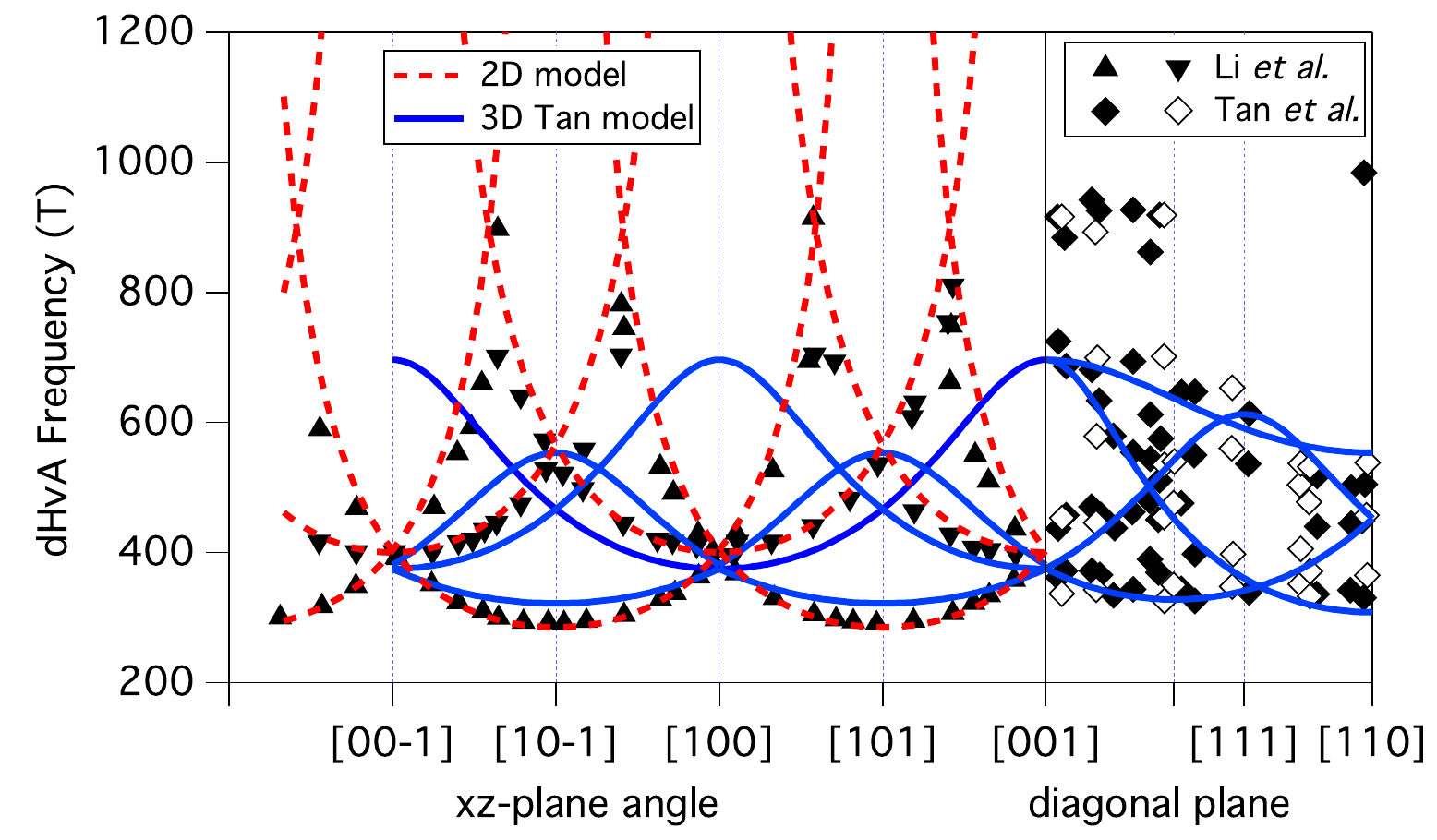}
\caption{(Color online)
Angular-dependence of the dominant amplitude dHvA frequencies in the 200-1000T range from two separate studies in different angular planes with fits to 2D and 3D models.   Dashed lines are a fit to \110b\ facet 2D cylindrical orbits \cite{Li14} with 285 T minimum cross section orbits.  Solid lines correspond to the anisotropic \110b-oriented 3D ellipsoid fit \cite{Tan15}   to the diagonal plane data with 300 (700) T minimum (maximum) frequencies extended to the $x$-$z$ plane data where significant disagreement is found.
}
\label{dhva1kT}
\end{center}
\end{figure}

Fig. \ref{dhva1kT} provides such an analysis for the $x$-$z$ plane data set in the 0.3-1 kT range.  It is important to note that the angle-to-angle connectivity of orbit frequencies in the $x$-$z$ plane data set is easily inferred from its wider angular range spanning multiple symmetric cubic axes with uniform angle spacing.   In contrast, model independent connectivity of frequencies is absent in the diagonal plane data sets that span a narrower angular range with irregular angle steps having up to 19\deg\ gaps.  Hence the $x$-$z$ plane data are very restrictive on model fitting.  As shown in Fig. \ref{dhva1kT} (dashed curves) and by Li \etal, the $x$-$z$-plane data are very nicely fit to $F_0$/cos($\theta$-$\theta_0$) with minimum frequencies of $F_0$=285 T at 45\deg\ and 135\deg\ and $F_0$=400 T at 90\deg\ and 180\deg, resulting in  prominent divergences every 45\deg.  This model corresponds to 2D cylindrical orbits oriented in multiple \110b directions \cite{suppl}.
The size of the 2D fit orbit corresponds to an area of 2.77 nm$^{-2}$ and a \kF\ radius of 0.094 \invA\ as reported previously \cite{Li14}.  
The 2D $x$-$z$ plane model also gives a very reasonable description of the combined diagonal plane data sets of Li \etal\ and Tan \etal\ \cite{Erten15,suppl}.

Can this $x$-$z$ plane data set possibly be compatible with a 3D orbit model? In fact, highly anisotropic ellipsoids elongated in \110b\ directions can fit the lower frequency portion of the data set equally well, but also predict, in the divergent regions of the 2D model, finite frequencies \cite{suppl} which are absent in the experimental data.  Decreasing anisotropy of a 3D ellipsoid progressively causes greater disagreement with the higher frequencies of the $x$-$z$ plane data.  This tension is very clear for the analysis of Tan \etal\, which posits \110b\ 3D ellipsoids taken to be analagous to the small intra-neck features of La\B6 \cite{suppl}.  As shown in the right and left panels of Fig. \ref{dhva1kT}, respectively, we have reproduced their diagonal plane analysis and then extended it to the $x$-$z$ plane.  There we see only a small discrepancy in the minimum frequencies of 300-350 T, but the maximum orbit frequency of 700 T is clearly inconsistent.  We note in passing that, in detail, this model is actually very inconsistent in orbit size (80 times larger) and ellipsoid anisotropy with what would be expected from the attempted analogy to the La\B6\ intra-neck orbits \cite{Harima88}, and that a model with anisotropy true to the La\B6\ analogy also fails badly for the $x$-$z$ plane data \cite{suppl}.  
Similarly, the even lower orbit frequency range of 30-150 T exhibits $x$-$z$ plane minima along $<$100$>$ directions \cite{Li14}, which are not consistent with \110b\ 3D intra-neck ellipsoids.
In fact, we show below that the reduced-band filling for Sm\B6\ would preclude the existence of the necks (if the material were actually metallic).


The most prominent and well established 2D surface state Fermi surface feature in previously reported ARPES for the cleaved (001) surface is much larger than found in the dHvA of Li \etal.  Thus the report of new dHvA with orbit frequencies $>$1 kT naturally invites comparison to ARPES.  Furthermore the 2D analysis of the 0.3-1 kT dHvA in Fig.  \ref{dhva1kT} and in Li \etal\ highlighting (110)-facet 2D orbits
motivates the search by ARPES for 2D Fermi surfaces also on the Sm\B6(110) surface.  We carried out ARPES studies on both (001) and (110) surfaces. The (001) surfaces were prepared by $in$-$situ$ cleaving  in ultra-high vacuum better than $8 \times 10^{-11}$ Torr. The (110) surface of Sm\B6\ was prepared by polishing of a Laue-oriented  single crystal followed by  in-vacuum ion sputtering and high temperature annealing to 1300$^\circ$C. Low energy electron diffraction confirmed the formation of  1$\times$1 rectangular  long-range surface order  with $\sqrt2$ aspect ratio. The ARPES measurements were performed at the  MERLIN Beamline 4.0.3 at the Advanced Light Source (ALS)  synchrotron, using a Scienta R8000 electron  energy analyzer and a low temperature 6-axis sample manipulator  cooled with an open-cycle He flow cryostat.


\begin{figure}[t]
\begin{center}
\includegraphics[width=8.5cm]{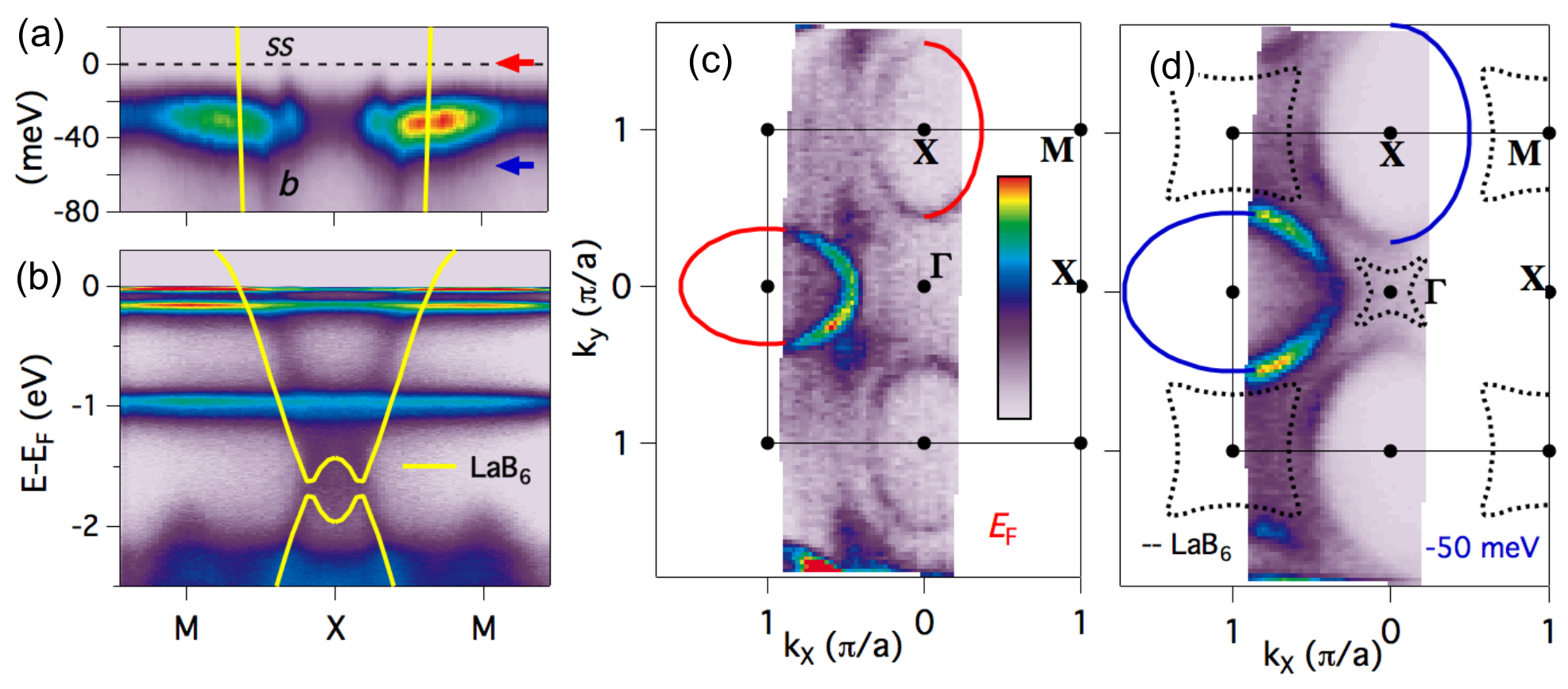}
\caption{(Color online)
ARPES measurements of cleaved Sm\B6(001) at T=6K. (a) M-X-M near-\EF\ spectrum showing in-gap states. (b) Wider spectrum showing the full bulk $d$-band electron pocket with comparison to La\B6\ DFT band structure.  (c) Fermi-edge intensity map showing quantification of the 2D in-gap state size. (d) Energy cut at -0.05 eV showing bulk $d$-band contours with comparison to the trivalent La\B6\ DFT Fermi contours. All measurements acquired at $h\nu$=70 eV.
 }
\label{arpes001}
\end{center}
\end{figure}

Figure \ref{arpes001} shows the basic ARPES results for the cleaved (001) surface of Sm\B6\ with a focus on quantifying  the X-point contour sizes of the 2D in-gap surface state and the bulk $d$-band electron pocket and comparing to the trivalent La\B6\ band structure. 
The presented ARPES data were obtained at 70 eV photon energy in a high symmetry constant energy plane that cuts through the zone center so that bulk ($\Gamma$, X, M) and surface  (\Gbar, \Xbar, \Mbar) $k$-points are equivalent.

Figures \ref{arpes001}(a) and (b) present energy dispersion cuts along M-X-M,  respectively, very close to \EF\ showing the presence of the in-gap states and in a 2 eV range showing the full $d$-band electron pocket dispersion. A low temperature (T=6 K) Fermi surface map of the in-gap states, shown in Fig. \ref{arpes001}(c), allows quantification of the X-point elliptical contour size of \kF=(0.27,0.39)$\pm$0.01 \invA, corresponding to an area of 33$\pm$2 nm$^{-2}$ and a dHvA orbit  frequency of 3.5$\pm$0.2 kT.   
A key result of this paper is that such a frequency is found in the diagonal scan data of Tan \etal\ along the [001] axis.  
The two-dimensionality of the in-gap states has previously been firmly established by ARPES \cite{Xu13,Jiang13,Min14,Denlinger13b}.

The ARPES data also demonstrate the very large and significant difference in the $d$-band fillings of trivalent La\B6\ and mixed valence Sm\B6.  Overplotted in panels (a) and (b) is the DFT-calculated band structure \cite{Wien2k} for trivalent La\B6.
The $d$-band X-point electron pockets for La\B6\ enclose $\approx$50\% of the bulk cubic BZ 
(with appropriate correction for the twelve \110b\ small $\approx$1\% overlap volumes),  corresponding  to 1 electron per unit cell.  In contrast the extrapolated $d$-band \EF\ crossings for Sm\B6\ imply only about 25\% of the BZ \cite{suppl} in full agreement with the well-established Sm\B6\ mixed-valence of +2.5, i.e. 0.5 electrons per unit cell. The in-gap surface state contours are even smaller. Panel (d) shows an angle dependent map for a binding energy just below the shallowest $f$-state (\mbox{-0.05} eV) and reveals bulk $d$-band contours that are larger than the in-gap state contours and only slightly smaller than the size upon extrapolation to \EF.  This contour has a size of $k_{||}$ = (0.38, 0.52) \invA, area=62 nm$^{-2}$, and F=6.5$\pm$0.3 kT.  The equivalent FS contour in La\B6\ has a frequency of 10 kT, significantly larger than any (001) contour observed in Sm\B6.   Also overplotted are calculated La\B6\ M-point star shaped FS contours which result from the X-point ellipsoidal overlap.  Such overlap features are precluded by the much smaller Sm\B6\ filling, and indeed, dHvA measurements show that 10\% alloying of Sm into La\B6\ is sufficient to reduce the X-point ellipsoid size to just remove the overlap neck \cite{Goodrich09}.


\begin{figure}[t]
\begin{center}
\includegraphics[width=8.5cm]{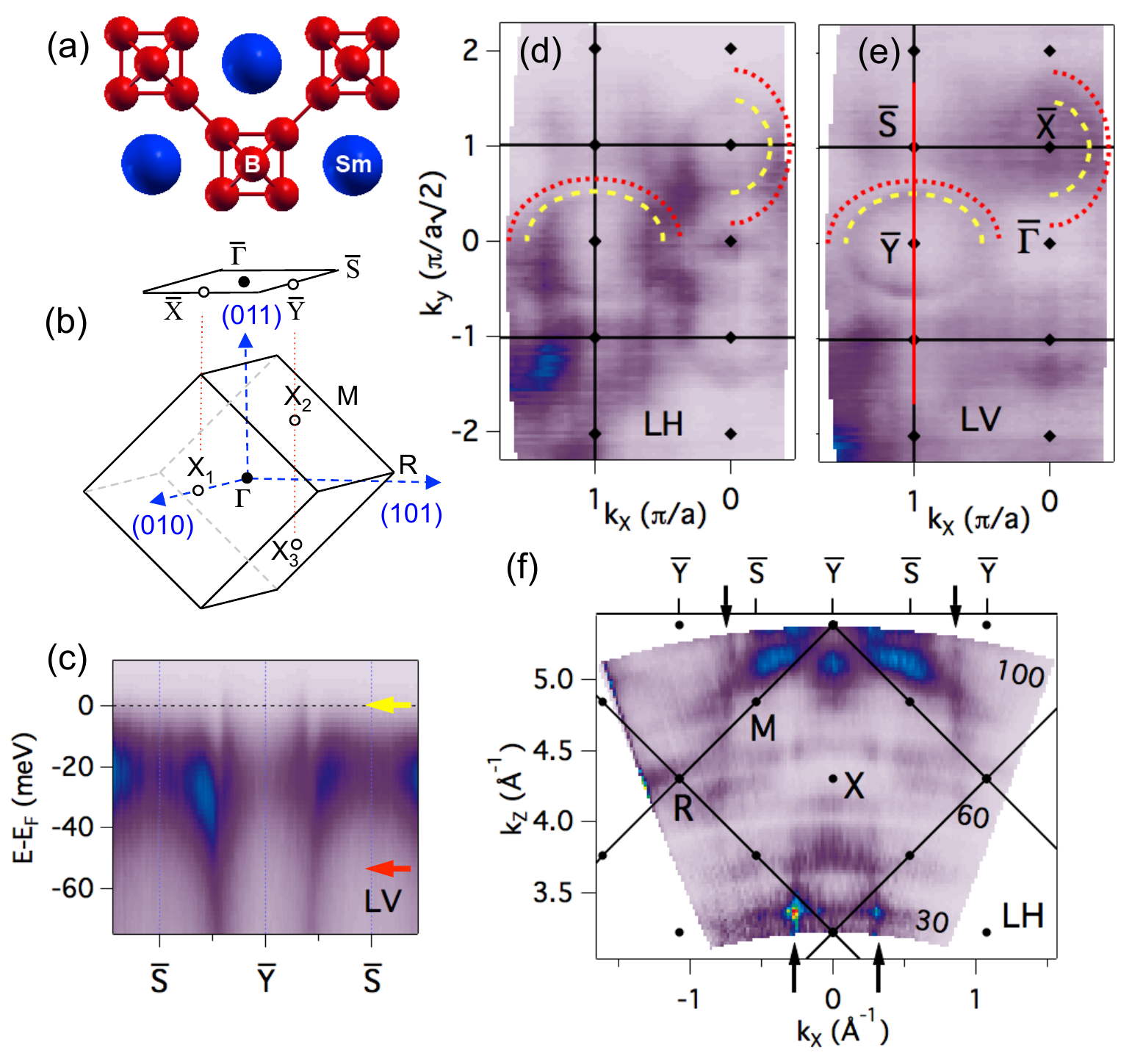}
\caption{(Color online)
ARPES measurements of prepared Sm\B6(110) at 15K. (a) Side view schematic of the hexaboride (110) surface. 
(b) Cubic bulk Brillouin-zone with (110) rotation and projection to the rectangular surface BZ. 
(c) Near-\EF\ band dispersion cut through the \Ybar-point for LV-polarization (red line in (e)) showing in-gap states similar to the (001) surface.
(d,e) Fermi surface maps using LH and LV-polarization of the incident 66 eV x-ray excitation.  Dotted line half-ellipses show quantification of the 2D in-gap state sizes (yellow) and -0.05 eV bulk contours (red, not shown).
(f) Photon-dependent $k_x$-$k_z$ \EF\ intensity map  at the BZ boundary $k_x$=0.76 \invA\ (\Ybar\ cut) using LH-polarization. Vertical streaks (arrows) indicate 2D non-bulk character.
 }
\label{arpes110}
\end{center}
\end{figure}

Figure \ref{arpes110}(a) shows a schematic of the hexaboride (110) surface in which it is observed that cations lie in-plane with \B6-octahedra. Hence the (110) surface is non-polar in contrast to the (001) surface which is highly polar with opposite sign charges for Sm- and \B6-terminations and leads to spatial variations of spectra from cleaved (001) surfaces \cite{Denlinger13,Hlawenka15}.
Figure \ref{arpes110}(b) shows the cubic hexaboride Brillouin zone rotated 45 degrees and with projection to the (110) surface Brillouin zone that is rectangular in shape with $\sqrt2$ aspect ratio.  It is observed that the bulk X$_1$ point projects onto \Xbar, whereas both X$_2$ and X$_3$ project onto \Ybar\  of the surface BZ.

Photon-dependent mapping in the 30-150 eV range along normal emission and at the \Xbar\ or \Ybar\ zone-boundaries was performed in two different azimuth angle orientations in order to establish the location of bulk X-points \cite{Kang15}, to investigate the two-dimensionality of in-gap states and to determine the best photon energies for detailed angular-dependent $k_x$-$k_y$ maps.  
A zone center $\Gamma$ point is found at $h\nu$=61 eV (inner potential $V_0$=14 eV) and $h\nu$=66 eV favorable for angular scan mapping. 
Figure \ref{arpes110}(c) shows a near-\EF\ spectrum along the \Ybar\ zone-boundary at 66 eV that is very similar to the (001) surface spectrum in Fig. \ref{arpes001}(a)  including the basic spectral ingredients of a bulk $d$-band electron pocket centered on \Ybar\ that disperses upwards (from -1.8 eV) and hybridizes with flat Sm 4$f$ states centered at -20 meV.  Then emerging from the $f$-states in the center of $d$-band pocket are metallic in-gap states that cross \EF\ at $\pm$0.14 \invA.

Fermi surface maps of these in-gap states are presented in Figs. \ref{arpes110}(d) and (e) using linear horizontal (LH) and linear vertical (LV) polarizations of the incident 66 eV x-rays to selectively highlight different FS features.  The LV map reveals better an elliptical contour at \Ybar, and so it is used to quantify the \kF\ radii, (0.38, 0.28)$\pm$0.02 \invA,  and the FS area ($\pi k_1 k_2$)=33$\pm$4 nm$^{-2}$.  Similarly the LH map reveals better the smaller contour at \Xbar\ which is approximated as an ellipse for quantitative size analysis of \kF=(0.28, 0.22)$\pm$0.02 \invA and  FS area of 19$\pm$3 nm$^{-2}$.

The predicted dHvA frequencies for the \Ybar\ and \Xbar\ FS ellipses are then computed from the Onsager relation to be 2.0$\pm$0.3 kT and 3.5$\pm$0.4 kT. 
The large \Ybar\ ellipse has a size and aspect ratio that is essentially the same as the two \Xbar\ in-gap states for the cleaved (001) surface as in Fig. \ref{arpes001}(c).  
Fig. \ref{arpes110}(e) shows a $k_x$-$k_z$ FS map where $k_z$ is accessed by varying the photon energy.  For fixed photon energy, angle variation gives a circular k-path so that variations of intensity with photon energy generates the appearance of circular arcs.  Significant here, however, are the vertical streaks that confirm the $k_z$ independent FS that is characteristic of 2D surface states.


\begin{figure}[t]
\begin{center}
\includegraphics[width=7.5cm]{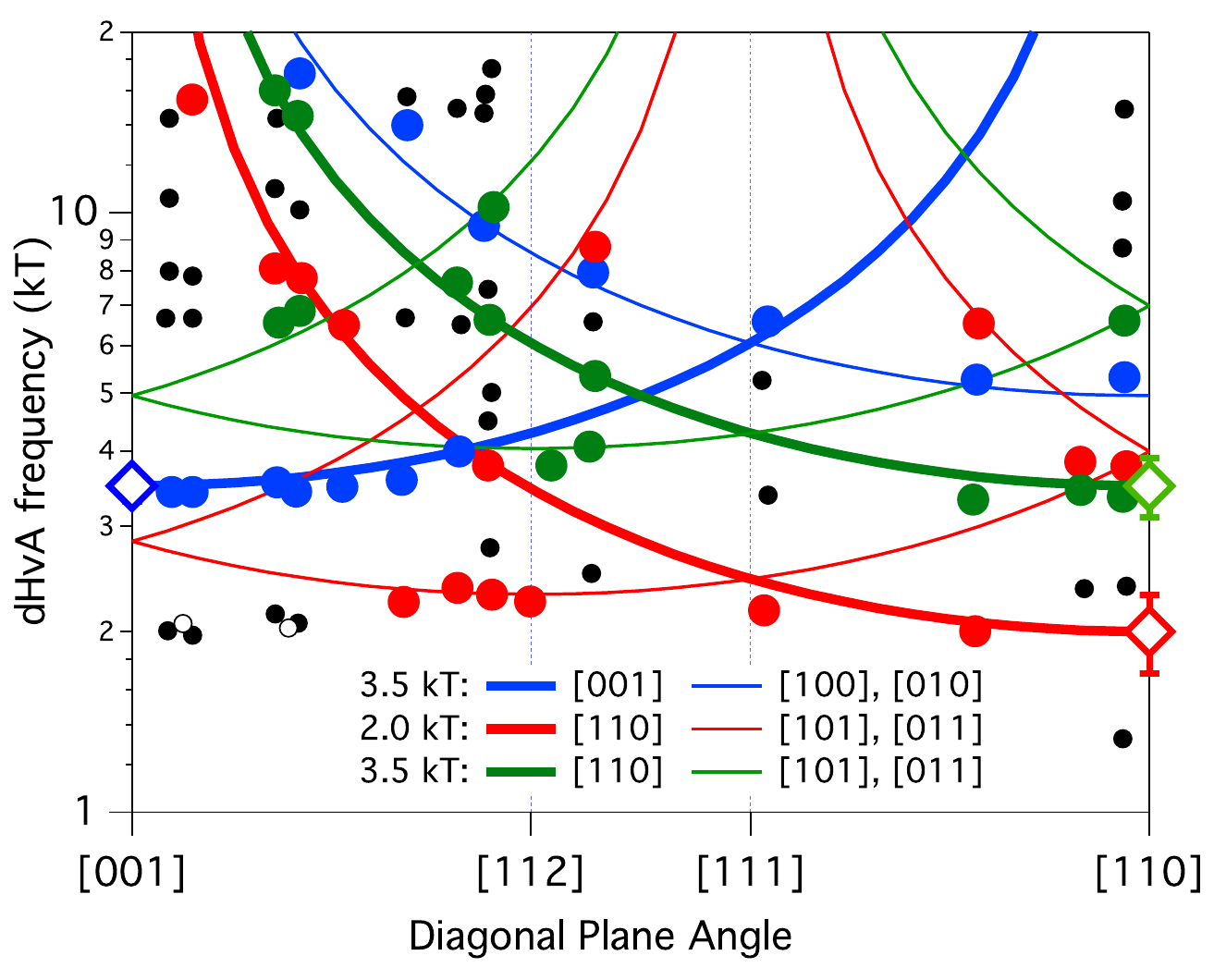}
\caption{(Color online)
(a) High frequency ($>$2 kT) dHvA orbit frequency diagonal-plane angular dependence from Tan \etal\  with a 2D model based on the ARPES in-gap state orbit sizes (diamond symbols) from the (001) and (110) surface measurements. Thin lines correspond to frequencies expected from other $<$001$>$ and \110b\ facets,
differing only by numerical factors coming from the geometry. Data points are color-coded to matching lines.
}
\label{dhva10kT}
\end{center}
\end{figure}

Next we compare in Fig. \ref{dhva10kT} the experimental ARPES results to the new high frequency dHvA data reported by Tan \etal.  
To begin, we note that the oscillation amplitude for these kT size pockets is at most only 10\% of that of the dominant oscillations \cite{suppl}.   It would be impossible to exclude the possibility of oscillations coming from small facets such as these $<$100$>$ and \110b\ family surfaces. 
The three diamond symbols correspond  to the in-gap surface state FS sizes from the (001) \Xbar\ and from the (110) \Xbar\ and \Ybar\ points.   Assuming two-dimensionality of the states as measured  by ARPES, three sets of inverse cosine dHvA branches appropriate for $<$001$>$ and \110b\ facets are then overlaid on the plot. 
Each of the three sets of color-coded branches connects a minimum of 15 data points, totaling more than half of all the dHvA frequencies.   Note that this is not the result of a fit and there are no free parameters.

As presented in the Supplement \cite{suppl}, additional data points can be accounted for by the inclusion of another set of branches for either a $<$001$>$ or a $<$111$>$ facet 2D surface state.
The success of the 2D model to account for so many dHva data points demonstrates that it is at the very least an equally plausible alternative to the horizontal connectivity of the angular data that is proposed in the 3D model of Tan \etal.   The divergences of the 2D branches can account for high frequencies without having to resort to large trivalent X-point ellipsoids that have no basis in the ARPES data, and are completely inconsistent with the mixed valence $d$-band-filling.

In conclusion we have provided multiple arguments to challenge the recent claims of the exotic existence of bulk trivalent-like 3D metallic orbits in the bulk insulating phase of mixed-valent Sm\B6.  ARPES near-\EF\ electronic structure is fully consistent with non-overlapping mixed-valent bulk $d$-band X-point pockets and provides no foundation for  trivalent La\B6-sized orbits.  Instead ARPES-measured in-gap surface state FS orbits on (001) and (110) surfaces of  Sm\B6\ agree well with orbit frequencies in the new $>$1 kT dHvA data and provides a parameter-free basis for a successful 2D model of its angular dependences.  Finally, the originally reported lower frequency $<$1 kT dHvA orbit angular dependence in the $x$-$z$ plane \cite{Li14} 
definitively favors a 2D model of (110) facet surface states.

\acknowledgments

The U.S. DOE supported ARPES studies at the Advanced Light Source (DE-AC02-05CH11231), and dHvA torque magnetometry studies at the University of Michigan (DE-SC0008110).   JWA acknowledges visitor support at the Advanced Light Source during the period when this paper was written.  KS was supported by the U.S. NSF (Grant No. ECCS-1307744).  


\end{document}


\def\B6{B$_{6}$}
\def\EF{$E_\textrm{F}$}
\def\kF{$k_\textrm{F}$}
\def\vF{$v_\textrm{F}$}
\def\invA{\AA$^{-1}$} 
\def\deg{$^\circ$}
\def\etal{$et\ al.$}
\def\Gbar{$\overline{\Gamma}$}
\def\Xbar{$\overline{\textrm{X}}$}
\def\Ybar{$\overline{\textrm{Y}}$}
\def\110b{$<$110$>$}
\def\cred{\color{red}}
\def\cblue{\color{blue}}

\title{{\it Supplementary Material:}\\
Consistency of Photoemission and Quantum Oscillations\\ for Surface States of SmB$_6$
}

\author {J. D. Denlinger}
\affiliation{Advanced Light Source, Lawrence Berkeley Laboratory, 
	Berkeley, CA 94720, USA}
\author {Sooyoung Jang}
\affiliation{Advanced Light Source, Lawrence Berkeley Laboratory, 
	Berkeley, CA 94720, USA}
\affiliation{Department of Physics, Pohang University of Science and Technology, 
	Pohang 37673, Korea}
\author {G. Li}
\author {L. Chen}
\author {B. J. Lawson}
\author {T. Asaba}
\author {C. Tinsman}
\author {F. Yu}
\author {Kai Sun}
\author {J. W. Allen}
\author {C. Kurdak}
\affiliation{Department of Physics, Randall Laboratory, 
	University of Michigan, Ann Arbor, MI 48109, USA}  
\author {Dae-Jeong Kim} 
\author {Z. Fisk}
\affiliation{Department of Physics and Astronomy, 
	University of California at Irvine, Irvine, CA 92697, USA}
\author {Lu Li}
\affiliation{Department of Physics, Randall Laboratory, 
	University of Michigan, Ann Arbor, MI 48109, USA} 

\date{\today}

\maketitle


\textbf{Contents} \\
1. Quantum Oscillation amplitudes\\
2. General Aspects of Trivalent Hexaboride dHvA\\
3. Michigan $x$-$z$ plane $<$1 kT dHvA data \\
4. 2D fit of Diagonal plane $<$1 kT dHvA data\\
5. Trivalent versus Mixed-Valent dHvA orbits\\
6.  Additional Diagonal plane $>$1 kT 2D dHvA assignments\\

\subsection{1. Quantum Oscillation amplitudes}

\begin{figure*}[t]
\includegraphics[width=14.5 cm]{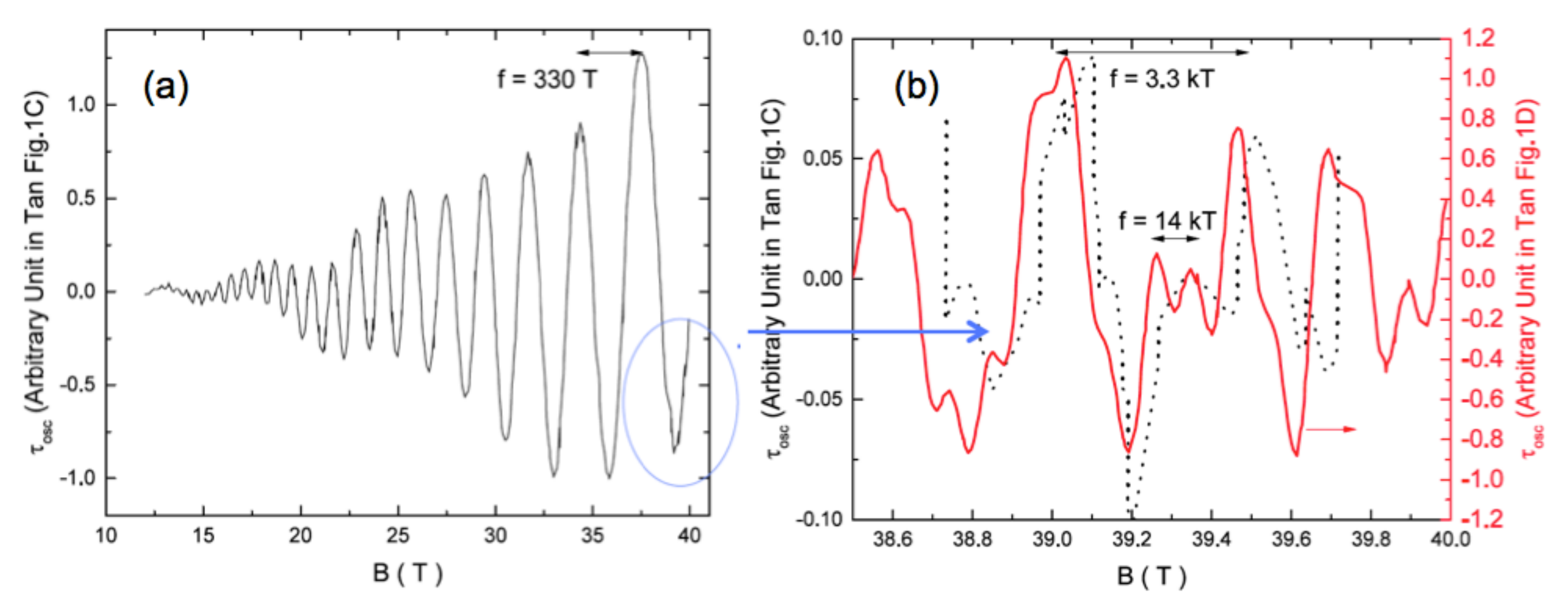}
\caption{
(a) Example background-subtracted magnetic torque dHvA oscillation from Tan \etal\ \cite{Tan15} showing the dominant 330 T oscillation with amplitude increasing to a maximum of $\pm$1 for B$>$35 T.   (b) Residual high-frequency oscillation signal for B$>$38 T after subtraction of a 330 T sinusoidal period revealing 3.3 kT and 14 kT oscillation periods with $\sim$10\% and $\sim$1\%  relative amplitudes. }
\label{dhvaAmpl}
\end{figure*}

Knowledge about the absolute or relative amplitude of the $>$1 kT frequencies is useful for informed discussion of the comparison of Sm\B6\ dHvA to that of the trivalent hexaborides.  While early published dHvA data on La\B6\ and Pr\B6\ \cite{Ishizawa77,Onuki85} 
did not show FFT spectra with relative amplitudes, more recent studies of Nd\B6\ \cite{Goodrich06} and La\B6\  \cite{Goodrich09}  show the dominant FFT amplitude for the 8-10 kT X-point $\alpha$-orbits.
In contrast, the dominant dHvA oscillation amplitude for Sm\B6\ in both dHvA studies is for the lower $\sim$300 T frequencies \cite{Li14,Tan15}.   
Fig. 2 of Tan \etal\ \cite{Tan15} presented three FFT spectra spanning three different frequency ranges derived from one angle direction using different field scan windows, but with arbitrary amplitude scales for each. An example wide field-range quantum oscillation signal was also shown in Fig. 1C of Tan \etal\ after subtraction of a background which is dominated by a 330 T frequency.
Further subtraction of the 330 T oscillation in a high-field region is also shown in Fig. 1D of Tan \etal\  to highlight the higher frequency oscillations, but with a separately normalized arbitrary unit amplitude scale.  Hence we have digitized the wide-field quantum oscillation data in Fig. \ref{dhvaAmpl}(a), and performed the same low frequency subtraction to obtain the residual higher frequency oscillations,  plotted in Fig. \ref{dhvaAmpl}(b) using the same original amplitude scale.  Also plotted is the Tan \etal\  subtracted high frequency oscillation profile for a consistency check of the angular pattern. 

Whereas the dominant 330 T period oscillation has a signal amplitude at 38 T of $\approx$ 2, the maximum amplitude of a 3.3 kT oscillation in the subtracted signal  is less than 0.2 in the same unit scale as the 330 T data.  Also a higher 14 kT frequency component has $\approx$10$\times$ smaller amplitude than that for 3.3 kT.  Hence we can estimate from this one representative angle,  that the magnetization oscillation amplitudes for the f=3.3 kT and f=14 kT signals are less than 10\% and 1\% of the dominant f=330 T oscillation signal, respectively.

\subsection{2. General Aspects of Trivalent Hexaboride dHvA }

Tan \etal\  provide a side-by-side comparison of their Sm\B6\ dHvA frequency angular dependences to those of Pr\B6\ and to La\B6\ as part of their argument for the interpretation and assignment of the Sm\B6\ orbits as being trivalent-like hexaboride orbits.  The very appealing idea is that the combination of the La\B6\ spectrum, which has orbit frequencies in well separated high and low ranges, with that of Pr\B6, which has many intermediate sized orbit frequencies, provides an overall visual appearance of general agreement with the complicated Sm\B6\ dHvA spectrum, including a transferrence of orbit labeling from mateial to material.  Unfortunately an examination of the dHvA trivalent hexaboride literature reveals serious conceptual problems with the presumption that there are such universal trivalent Fermi surface features between La\B6\ and antiferromagnetic Pr\B6.  A full review of the literature is inappropriate here, but we point out below the essence of the problems.

\textbf{La\B6.}
    The La\B6\ dHvA angular dependence is very well understood \cite{Ishizawa77,Onuki89ce} in terms of the large 8-10 kT $\alpha$ X-point electron ellipsoids that just slightly overlap each other forming necks that cause the $\alpha_{1,2}$ orbits to be absent along (001) in favor of smaller $\epsilon$ and $\gamma$ star-shaped hole orbits that complementarily exist only close to [001]. 
    The small overlap of the large X-point ellipsoids also produces very small anisotropic \110b-oriented ellipsoids within the necks with very small $<$10T $\rho$ orbits \cite{Harima88}.  In the other extreme, the very large $>$10 kT $\psi$ and $\lambda$ orbits exist over a short angular range due to orbits that connect through two Brillouin zones via the necks in a "figure eight" fashion \cite{Ishizawa77}.
   In contrast to La\B6, Sm\B6\ has many extra orbit frequencies spanning the full angular range in the 50-100 T, 0.4-1 kT, and 2-5 kT ranges.    Hence the pair-wise comparison of Sm\B6\ dHvA to that of La\B6\ alone is very weak in making a 3D trivalent orbit argument.  

\textbf{Pr\B6.}

The dHvA angular dependence of trivalent Pr\B6\ does contain orbit frequency bands that span the full angle range \cite{Onuki89ce}, apparently improving the comparison.  However, each of those frequency bands in Pr\B6\ has its own complex origin that arises from the Pr\B6\ anti-ferromagnetic (AF) electronic structure and the associated  zone-folding of the simpler overlapping X-point ellipsoidal electron FS into much more complicated hole and electron FS sheet topologies.  For instance, the band of  3-6 kT orbits  that span the full angle range originates from this AF zone-folding of large ellipsoids and hence are inappropriately labeled by Tan \etal\ as $\gamma$, since they have no connection to the La\B6\ $\gamma$ neck-induced hole orbit, which exists only close to [001].
 
      The smaller frequency Pr\B6\ orbits called $\rho$ (0.5-1 kT)  and $\rho'$ (100-200 T) have been assigned  to be AF exchange-split flat intra-neck ellipsoids  \cite{Onuki89ce}, based on the similarity of the angle dependence to the La\B6 $\rho$ orbits, but with the noted puzzle of the sizes being $\approx$30$\times$ and $\approx$110$\times$ larger in size than the tiny La\B6\ ellipsoids and more than twice the size of the necks implied by overlap of 8-10 kT X-point ellipsoids.    
A complex solution for the  $\rho$ and $\rho'$ orbits was proposed by Kasuya \etal\ \cite{Kasuya96}, involving a strong field dependence of the exchange-splitting that distorts the shape of the ellipsoids and a bonding-anti-bonding crossover effect that inverts the spin-up and spin-down ellipsoid sizes. 
How well-defined $\rho$ and $\rho'$ FFT orbit frequencies and angle dependences arise from the dHvA measurements with such steep field dependent distortions of size and shape is not explained. 
It is significant that no modern theoretical attempt at computing the AF Fermi surface has been published, perhaps due to the complex AF ordering vector of $Q=(1/4,1/4,1/2)$ below 7.0K involving 32 unit cells, as well as a spin-incommensurate magnetic structure below 4.2K \cite{Onuki85}.

\begin{figure*}[t]
\includegraphics[width=14.5 cm]{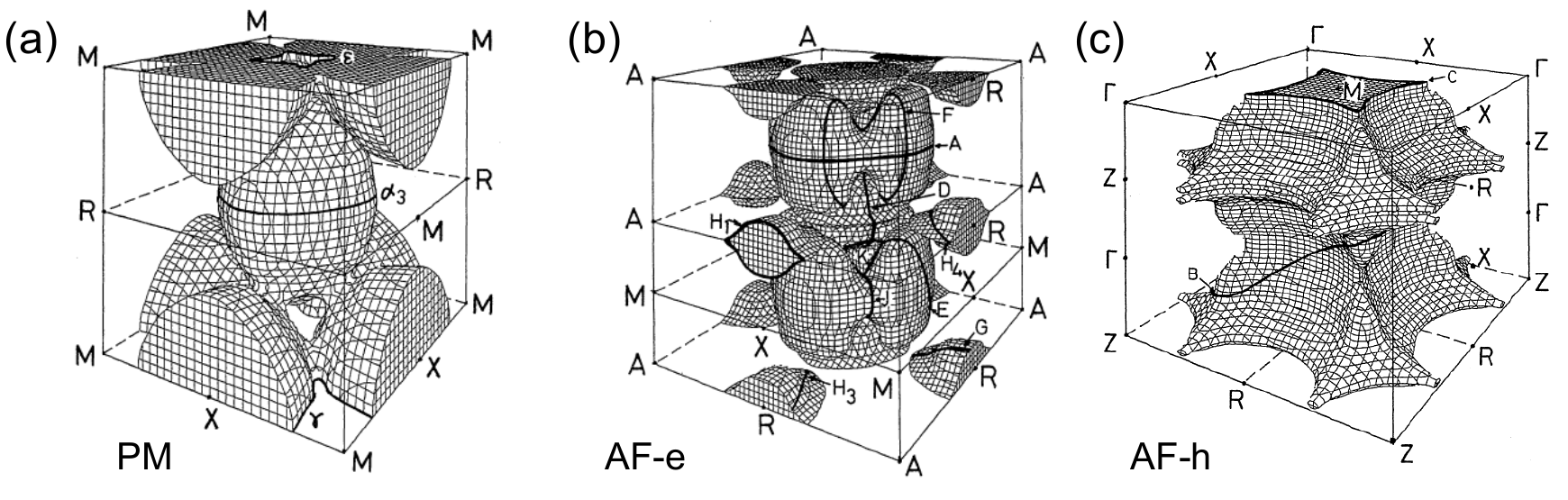}
\caption{
Comparison of (a) the La\B6-like PM phase electron Fermi surface of Nd\B6\ to the complex zone-folded (b) electron and (c) hole FS sheets of the AF phase. Reproduced from  Kubo \etal\ \cite{Kubo93}. 
}
\label{ndb6}
\end{figure*}

 \textbf{Nd\B6.}      
Similar to Pr\B6, trivalent Nd\B6\ is  antiferromagnetic below 7.8K, but with a simpler tetragonal order vector of $Q$=(0,0,1/2) involving a doubling of the unit cell.  We include Nd\B6\ here because there exist theoretical FS calculations \cite{Kubo93} that are useful for visualizing the issues pointed out above for Pr\B6.   Fig. \ref{ndb6} (a) shows the simple FS of the paramagnetic (PM) phase, essentially the same as for La\B6.    Fig. \ref{ndb6} (b,c) shows the complex electron and hole FS sheets of the AF phase, which yielded specific agreement to the angular dependence of $\approx$1 kT dHvA orbits.   By analogy, the FS topology of AF Pr\B6\ with a different AF ordering vector should be at least as complex as that of Nd\B6. 

We remark that the similarity of the PM phase FS to that of La\B6, even though Nd\B6\ has an even number of electrons per unit cell, arises because Nd is pure trivalent so that its 4$f$ states are well removed  by the large 4$f$ Coulomb interaction from the vicinity of the Fermi energy.  In contrast, for even electron but mixed valent Sm\B6, the 4$f$ states participate intimately in the near \EF\ electronic structure (to make gaps).  For Nd\B6, the La\B6-like orbits can be observed at the low temperatures of the dHvA measurements for B$>$27 T, upon which there is a first order transition from the AF FS topology to a spin-split paramagnetic-like Fermi surface \cite {Goodrich06}.  

The above brief review of the AF hexaboride dHvA literature reveals that the $\rho'$, $\rho$ and $\gamma$ frequencies are not ``universal Fermi surface features identified from experiment and band structure calculations,'' as claimed by Tan \etal.   Also none of the above complex AF zone-folding and exchange-split FS neck ellipsoids with strong field-dependence as proposed for Pr\B6\ (or Nd\B6) is likely applicable to the physics of mixed valent insulating Sm\B6.   Hence the $\rho'$, $\rho$ and $\gamma$ labeling of Sm\B6\ orbits to  suggest a common origin with that of Pr\B6\ orbits is inappropriate.

\subsection{3. Michigan $x$-$z$ plane $<$1 kT dHvA data}

Tan \etal\  present a dHvA data set for a 90 degree range along the diagonal rotation direction (001)-(111)-(110) , with FFT frequency analysis up to 20 kT.  Data for this angle range and direction were also presented in the supplement of Li \etal \cite{Li14} with analysis up to 1 kT.     In the $<$1 kT orbit range these two diagonal scan data sets can be directly compared to each other and to the proposed 2D and 3D FS orbit models.  Such a comparison has been made, with the conclusion that the two sets are compatible, but that, given the error bars, it is not possible to distinguish if a 2D or a 3D model fits better \cite{Erten15}.

\begin{figure*}[t]
\includegraphics[width=15.5 cm]{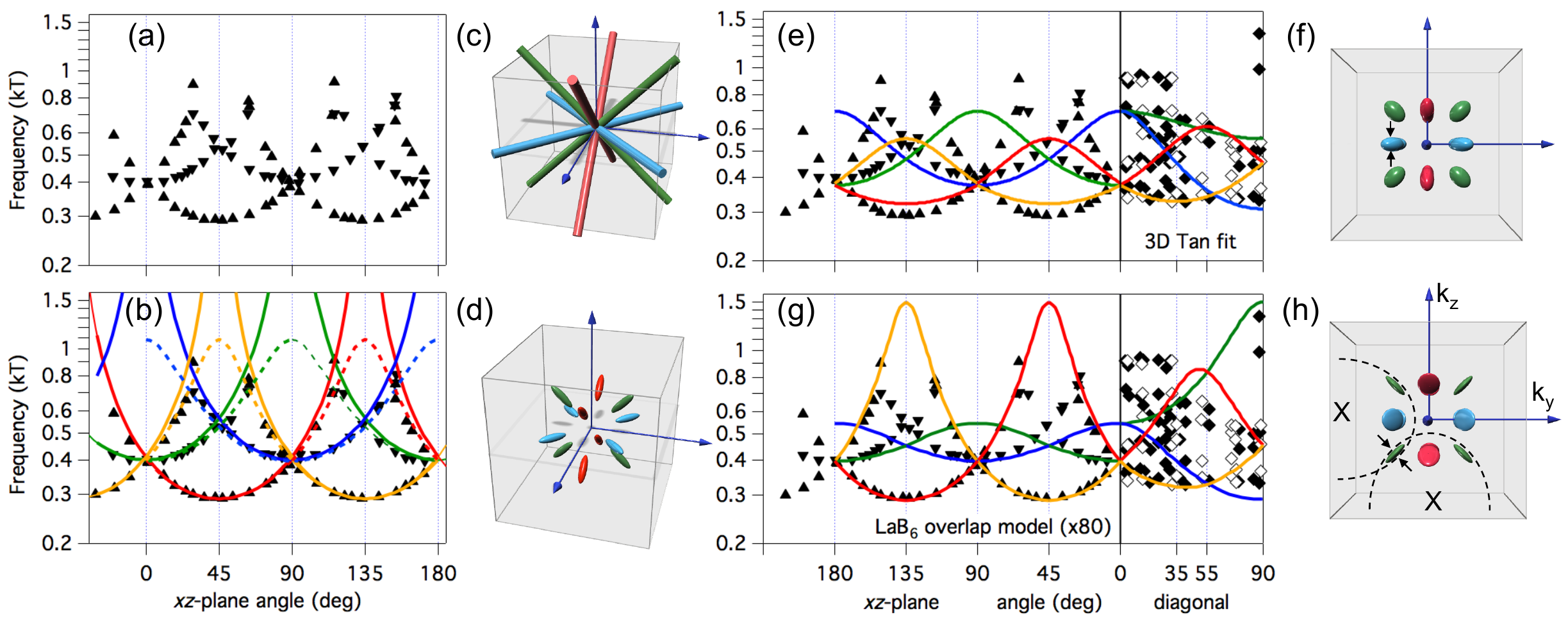}
\caption{
{\bf
Model fits to Michigan intermediate frequency dHvA angular dependent data of Sm\B6. } 
(a) No model interpretation. (b,c) 2D \110b-oriented cylinders. 
(b, d) Highly anisotropic \110b-oriented 3D ellipsoids (dashed lines). 
(e, f) Weakly anisotropic \110b\ ellipsoid model from Tan \etal\ \cite{Tan15}. 
(g.h) Highly flattened oblate ellipsoid model appropriate for large trivalent X-point ellipsoid overlap \cite{Harima88}. 
Fermi surface schematics are shown in (b,d,f,h) with perspective of front face views.  
Only the 2D  model can properly describe the dHvA angular dependences. 
}
\label{dhva300Tnew}
\end{figure*}

Absent from this comparison and conclusion is discussion of the main $x$-$z$ plane angular dependent data set of Li \etal\ covering 210 degrees spanning (001)-(011)-(010)-(01$\overline{1}$)-(00$\overline{1}$) and analyzed up to 1 kT.  That data set was presented by Li et al. in separate FFT plots for orbits with (110) and (001) minima.  Fig. \ref{dhva300Tnew}(a) presents the 1st harmonic FFT frequencies of the complete data set, but without any model fit.  It is clear from this ''raw'' data that the angle-to-angle connectivity of the orbit frequencies is easily inferred without a model (up versus down triangle symbols), in part due to the wide angular range allowing symmetric variations to be easily detected.  Such model independent connectivity of frequencies is absent in the current diagonal scan data sets.  As described in the main text, Fig. \ref{dhva300Tnew}(b,c) shows the 2D model of  Li \etal\ and the resulting excellent fit to the data.  Shortening the 2D cylinder Fermi surfaces to  highly elongated 3D ellipsoids, as shown in panel (d),  will equally well match the lower frequency angular dependence but, as shown also in panel (b), with the inherent prediction of finite higher frequency ellipsoid orbit closures every 45\deg, which are not seen in the data.  The large $>$3.5 aspect ratio of the cylindrically symmetric ellipsoids, with \kF\ values of (0.35, 0.094) \invA, required to account for the highest experimental frequencies above 800 T, provides a lower-bound to the anisotropy of 3D \110b\ ellipsoids.  Any smaller anisotropy of the 3D ellipsoids  will significantly disagree with the $x$-$z$ plane data.  Fig. \ref{dhva300Tnew}(e) shows both the diagonal scan data of Tan \etal\ and the $x$-$z$ plane data set of Li \etal.   Panel (f) shows the ellipsoidal model of Tan \etal.    Using this model, our best fit of the diagonal scan data, essentially the same as that of Tan \etal, is shown in panel (e) along with the prediction of this model for the $x$-$z$ plane.  As explained already in the main text, this model notably has a maximum frequency of only 700 T, which is entirely unable to account for the higher frequencies in the data.  

 The physical motivation of the model of Tan \etal\ is the small ellipsoids of the FS of La\B6, that exist within the necks of the overlapping trivalent-sized X-point ellipsoids \cite{Harima88}. The Sm\B6 fit ellipsoids, characterized by \kF=(0.19, 0.112, 0.89) \invA, depart substantially from the physical concept.  First, they are roughly 80 times larger than in La\B6.   Second,  we note in panel (f) that the maximum frequency upper branch in the diagonal plane decreases by $\sim$20\% from [001] to [110], which arises from the non-cylindrical symmety of the ellipsoids with $\sim$20\% smaller \kF\ in the $z$-direction.  Such flattening of \110b-oriented ellipsoids along cubic axes is highly inconsistent with the anisotropy appropriate for the small ellipsoids of  La\B6.   These overlap-neck ellipsoids are predicted to be highly-flattened perpendicular to the neck as illustrated in Fig, \ref{dhva300Tnew}(h), and give an opposite increasing angular dependence of the upper branch in the diagonal plane \cite{Harima88}.  If we use such a model, true to the actual asymmetry in La\B6, although scaled up in size as needed, the result is as shown in panel (g).  The highly oblate anisotropy actually gives a good 2D-like fit to the high and low frequencies of the $x$-$z$ plane angular dependence,  but with the previously noted prediction of orbit closures along \110b\ directions that are experimentally absent.   The highly oblate La\B6\ ellipsoid model also most significantly fails to describe the 2D-like divergences of the frequency branches with  $<$001$>$ minima. 

Hence we conclude from this exercise that the \110b-oriented 2D surface state scenario is indeed the best fit model for the dominant amplitude 0.3-1.0 kT FS orbits, and that the data are definitively incompatible with low aspect ratio 3D bulk ellipsoids such as proposed by Tan \etal, as well as the highly oblate $\rho$ ellipsoids actually found in La\B6.

\subsection{4. 2D fit of Diagonal plane $<$1 kT dHvA data}

\begin{figure*}[t]
\includegraphics[width=8.5 cm]{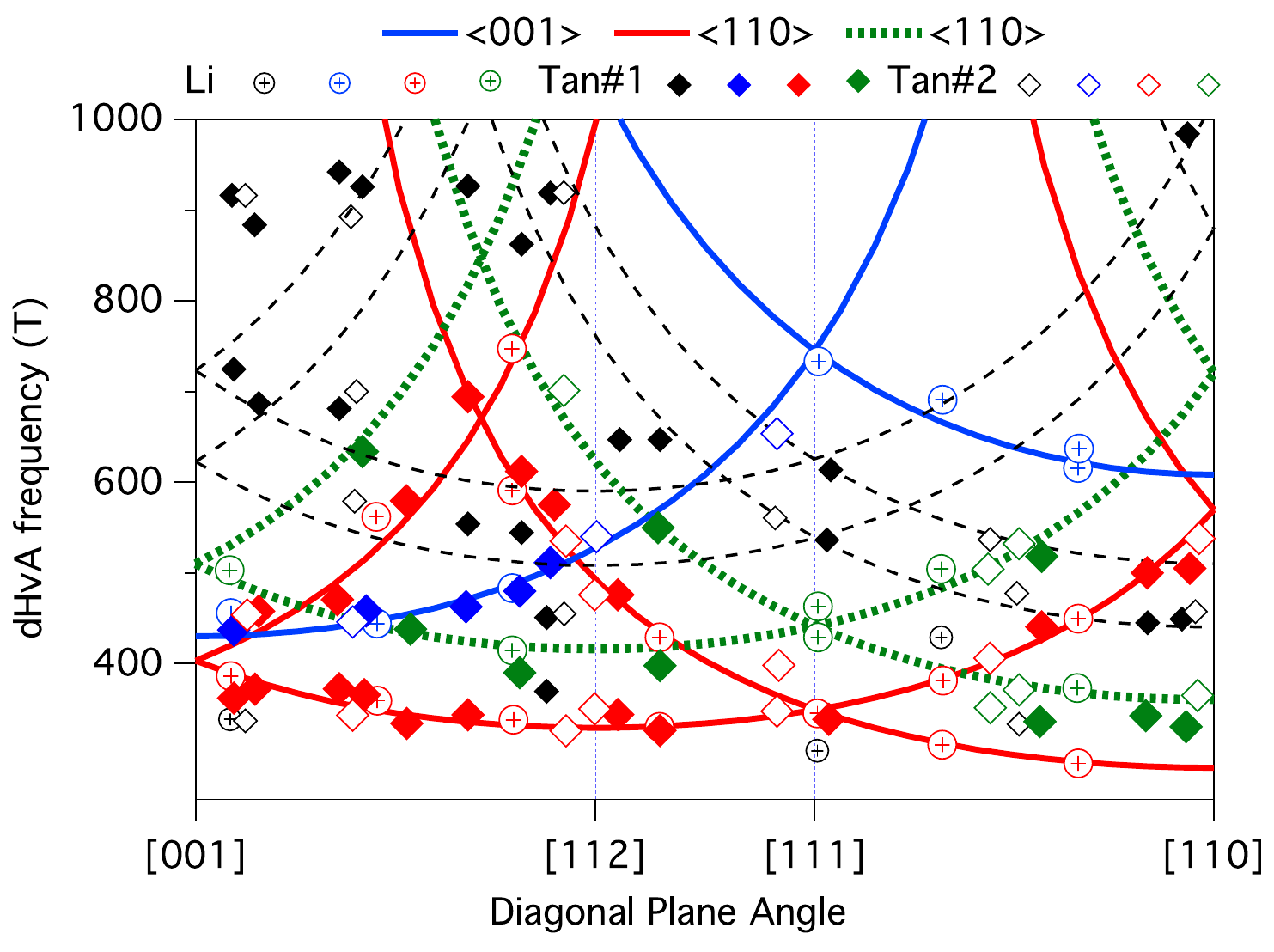}
\caption{
2D model analysis of the combined diagonal plane data sets of Li \etal\ and Tan \etal.   Solid lines represent primary \110b\ and $<$001$>$ facet 2D orbits. Color-coding of matching data points then reveal patterns of remaining data points that are suggestive of additional larger scaled \110b\ facet 2D orbits (dashed lines).   
}
\label{sfig_diag2d}
\end{figure*}

In this section we present a 2D model analysis of the combined 0.3-1 kT diagonal plane data sets from Li \etal\ and Tan \etal.  Two aspects are addressed: (i) the relative agreement between the data sets from the two groups, and (ii) the ability of 2D model orbits to explain the plethora of orbit frequencies which by eye do not reveal patterns of angle-to-angle connectivity as pointed out previously.  A plot of the combined diagonal plane data sets with the 2D model from Li \etal\ has previously been presented \cite{Erten15} with the conclusion that the data sets agree but that neither the 2D model nor the 3D model is preferred.
   Fig. \ref{sfig_diag2d} presents a similar plot but with an additional color-coding of data-points with sets of 2D orbits to which they match, and black data points representing unaccounted for frequencies.  The systematic methodology for arriving at the lines in Fig. \ref{sfig_diag2d} is as follows:

   (i) The \110b\ model fit to the $x$-$z$-plane data of Li \etal\ with $F_0$=285 T (from the previous section and Fig. 1 of the main text) is first plotted (red solid lines) and observed to nicely match $>$35 data points (approximately 1/3 of the $>$100 total number of frequencies) with no scaling adjustment necessary. Here we note that the data points matching these branches are not preferential to one data set or the other, attesting to the consistency of the data from the two groups.   Also we emphasize that the angular dependences of individual branches within a facet family cannot be independently scaled.

   (ii) Next we note that while the single \110b\ facet family of curves does not account for all the data points, the red color-coding of the matching data points reveals patterns in the remaining unaccounted for data points.  The first additional recognizable pattern is for a (001) surface 2D orbit and its companion (100) and (010) facets with $F_0$=430 T (blue solid lines), accounting for more than a dozen data points including the $\approx$700 T frequencies between [111] and [110]. Also we note that the near coincident minimum frequency of these $<$001$>$ 2D orbits  with that of the  \110b\ frequency along [001] means that in the $x$-$z$-plane, the angular dependences of these $<$001$>$ 2D orbits will overlap with those of the other (101) and (011) set of \110b\ facet angular dependences. At this stage we have reproduced the Li \etal\ supplemental section analysis.

   (iii) After blue color-coding of the $<$001$>$ branches, the remaining data points then strongly suggest the presence of a secondary \110b\ family of 2D orbits but scaled larger with $F_0$=360 T (green dashed lines).  This accounts for an additional dozen data points.  At this stage we note that all but two points of the Li \etal\ diagonal plane data set are accounted for, whereas the Tan \etal\ data set has a few dozen additional data points.
   
   (iv) Finally, elimination of the secondary \110b\  (green) data points then suggests two more parallel scaled \110b\ branches with [110] minima of 440 T and 510 T that accounts for most of the remaining data points.
   
    It could be argued from the final result of Fig. \ref{sfig_diag2d} that such a grid of orbits could explain any set of arbitrary diagonal plane data points.  However, the methodology for arriving at Fig. \ref{sfig_diag2d}, beginning as it does with the clear 2D fit of the detailed $x$-$z$-plane data set, provides a plausible 2D alternative to the 3D model proposed by Tan \etal, which is inconsistent with the detailed $x$-$z$-plane data set, and has, relative to its trivalent La\B6-like motivation, other deficiencies as discussed in the main text and in the preceding and following sections of the supplement.

\subsection{5. Trivalent versus Mixed-Valent dHvA orbits} 

As argued in the main text, ARPES of Sm\B6\ does not support the existence of trivalent-like large X-point ellipsoidal FS sizes giving rise to 8-10 kT dHvA orbit frequencies.  The following exercise quantifies what dHvA frequencies are to be expected from the ARPES-based electronic structure in relation to the trivalent hexaborides.

\begin{figure*}[t]
\includegraphics[width=14.5 cm]{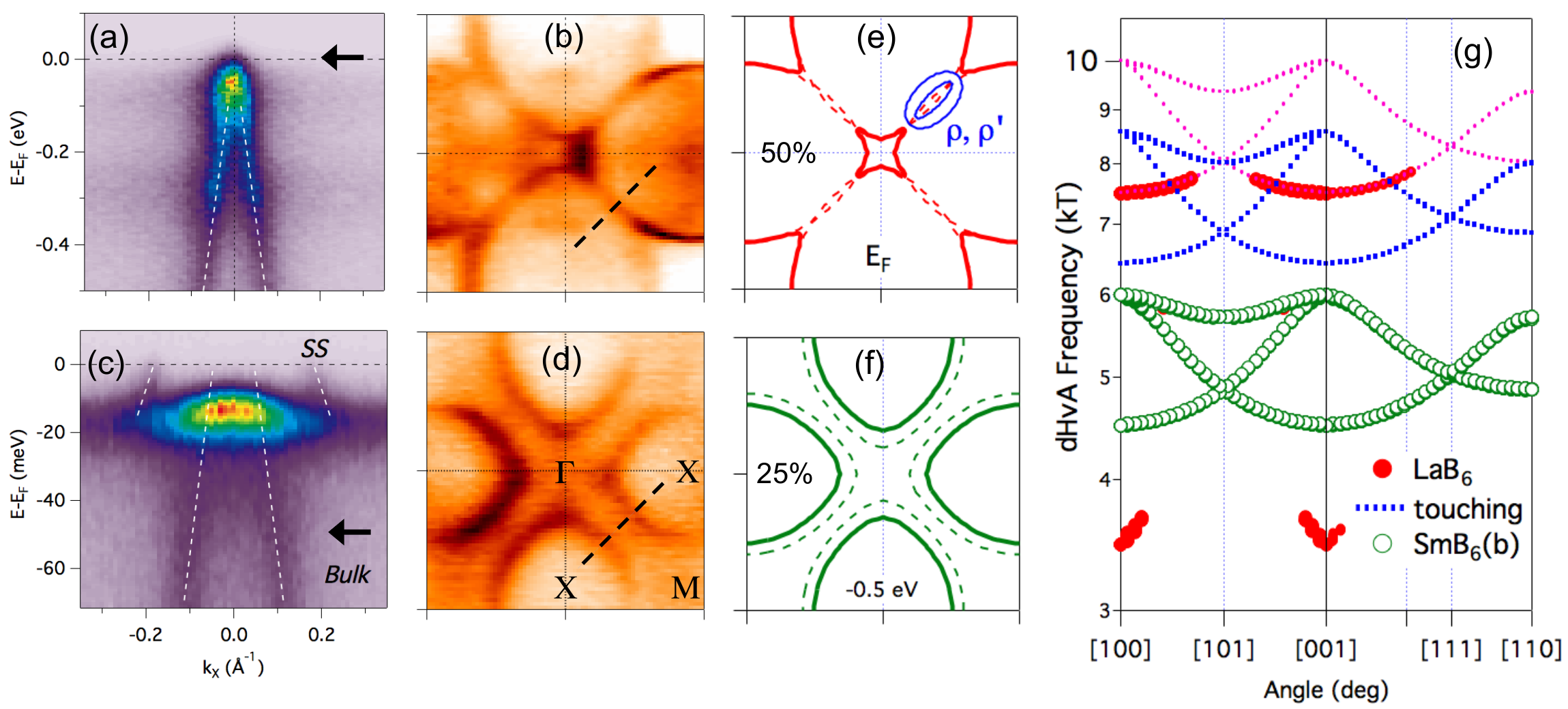}
\caption{
 (a-d) Comparison of h$\nu$=70 eV ARPES near-\EF\ electronic structure of (a,b) trivalent Nd\B6\ and (c,d) mixed-valent Sm\B6. (a,c) Energy dispersion  cut orthogonal through the closest approach of X-point ellipsoids (dashed lines in (b,d)).  Note energy scale difference.  (e,f) DFT energy contours of La\B6\ at \EF\ and at -0.5 eV selected to match the Sm\B6\ bulk contour size at -60 meV (arrow in (c)).  Dashed contour represents an extrapolation of the bulk Sm\B6\ $d$-band to \EF.  Blue contours represent the size of $\rho$ and $\rho'$ $<$110$>$ ellipsoid fits from Tan \etal\ \cite{Tan15}. (g) Comparison of predicted dHvA frequency angular dependences for the overlapping trivalent FS of (e), the mixed-valent FS of (f) and  an intermediate contour where the contours just touch.   
}
\label{dhvaDFT}
\end{figure*}

Fig. \ref{dhvaDFT}(a-d) provides a comparison of the FS of trivalent Nd\B6(001) to low T bulk $d$-band contours of Sm\B6(001) measured at -60 meV below \EF\ (just below the flat $f$-state).  
As explained already above in the general discussion of trivalent hexaborides, the FS of Nd\B6\ in the PM phase is essentialy the same as for La\B6.  The Nd\B6\ FS contours of \ref{dhvaDFT}(b) were measured in a high-symmetry plane at h$\nu$=70 eV and are observed to slightly overlap with each other.  A band dispersion cut orthogonal through this overlap region in \ref{dhvaDFT}(a), shows two $d$-band dispersions (\vF=5 eV-\AA) touching each other below \EF\ and forming a high spectral intensity point at $\approx$-50 meV. In contrast, the similar $k$-cut for Sm\B6\ shows the two $d$-band dispersions still separated from each other by $\approx$ 0.15\invA\ when they intercept the $f$-state at -20 meV.  Hence a constant energy map at -60 meV just below the $f$-state shows in Fig. \ref{dhvaDFT}(d) non-overlapping X-point elliptical-like contours. 
Extrapolations of these bulk $d$-bands with band velocity \vF=1.5 eV-\AA\ through the gap to \EF\ will still be separated and not produce a FS contour overlap.  

Fig. \ref{dhvaDFT}(c) also shows the presence of the 2D in-gap surface states and a spectral intensity ``hotspot'' and high energy dispersion of the $f$-band caused by the hybridization with the two $d$-bands. The T-dependence of this special non-overlapping ``hotspot'', labeled the ``$H$''-point, has previously been studied in depth \cite{Denlinger13b}.  At high T, the $f$-$d$ hybridization becomes incoherent and the bulk $d$-band disperses directly to \EF, similarly producing a non-overlapping FS at room temperature.

Next we quantify the bulk FS volumes of Nd\B6\ and Sm\B6\ by matching different energy-level contours of a DFT calculation \cite{wien2k} for trivalent La\B6\ to the experimental ARPES maps. 
The La\B6\ \EF\ contour size in Fig. \ref{dhvaDFT}(e)  is observed to be in very good agreement to that of the Nd\B6\ FS map in Fig. \ref{dhvaDFT}(b) including the small amount of overlap along the diagonal \110b\ directions ($\Gamma$-M).   This is expected since both are trivalent with their $f$-states sufficiently removed in energy 1 eV above or 6 eV below \EF,  for La\B6\ and Nd\B6, respectively.  Analysis of the enclosed volume of this 3D FS, accounting for the overlap, gives a fractional BZ filling of $\approx$ 50\% consistent with the trivalent one electron occupation of the $d$-band. 

Tuning the energy-level in the La\B6\ band calculation to -0.5 eV allows the size of the $d$-band contour to match the observed non-overlapping experimental -60 meV bulk contours of Sm\B6.  The large value of the energy shift is due to the light effective mass (m*$<$1) of the La\B6\ $d$-band. 
The fractional  FS volume filling of this contour, simply estimated by three ellipsoids of volume $\pi k_1^2k_2$ is $\approx$ 25\% of the bulk BZ, consistent with the well-established Sm\B6\ mixed valency of $\approx$ +2.5. A slightly smaller shift to -0.4 eV is required  to match an extrapolation of the $d$-band contour through the gap to \EF, with 28\% BZ filling (+2.55 valence).

Next the dHvA angular-dependent frequencies of the large X-point electron pocket are calculated from the theoretical La\B6\ 3D energy surfaces associated with the contours in Fig. \ref{dhvaDFT}(e,f) using a super-cell $k$-space extremal-area finder (SKEAF) program \cite{SKEAF}.

Consistent with the experimental dHvA of La\B6\ \cite{Ishizawa77}, the effect of the overlapping FS contours fragments the angular branches into finite segments including some higher and lower frequencies (not shown).
Fig. \ref{dhvaDFT}(g) shows the predicted dHvA angular-dependent frequencies for the 3D energy surfaces, giving the contours in Fig. \ref{dhvaDFT}(e) and (f).  
The angular dependence of non-interacting trivalent-sized ellipsoids, with 10 kT maximum frequency,  is also shown in Fig. \ref{dhvaDFT}(g) by scaling a smaller-sized non-overlapping ellipsoidal FS to match the 8 kT La\B6\ angular fragments. 

In contrast the non-overlapping dHvA frequencies for experimental Sm\B6, bulk contours are predicted to be in the 4.5-6 kT range, significantly less than for La\B6.  
If the applied magnetic field were to cause a breakdown of the hybridization gap and allow the bulk $d$-bands to disperse to \EF, the resulting non-overlapping FS and dHvA orbit sizes would be slightly larger than the -60 meV energy cut of Sm\B6.   There is no evidence reported yet of any field-dependent transition in the Sm\B6\ dHvA experiments.  

Tuning of the energy cuts within a DFT calculation of Sm\B6\  is also discussed by Tan \etal\  in order to speculate about the possible existence of trivalent-sized orbits in Sm\B6.  Their illustrated graphic example, however, is actually of non-overlapping ellipsoids that have tiny necks connecting  adjacent ellipsoids. This scenario of almost-touching ellipsoids is also plotted in Fig. \ref{dhvaDFT}(g), giving a dHvA frequency range of 6.5-8.5 kT.  Additional energy tuning within Sm\B6\ DFT calculations can produce overlapping contours but typically in regions just below the $f$-states with significant $f$-$d$ hybridization or far above \EF, and thus far from the reality of the ARPES-based electronic structure. 

This supplemental exercise provides additional graphical illustration of why the ARPES-measured electronic structure of Sm\B6\ does not support the existence of trivalent-sized dHvA orbits, implying that alternative explanations must be considered.

\subsection{6. Additional Diagonal plane $>$1 kT 2D dHvA assignments }

\begin{figure*}[b]
\includegraphics[width=14.5 cm]{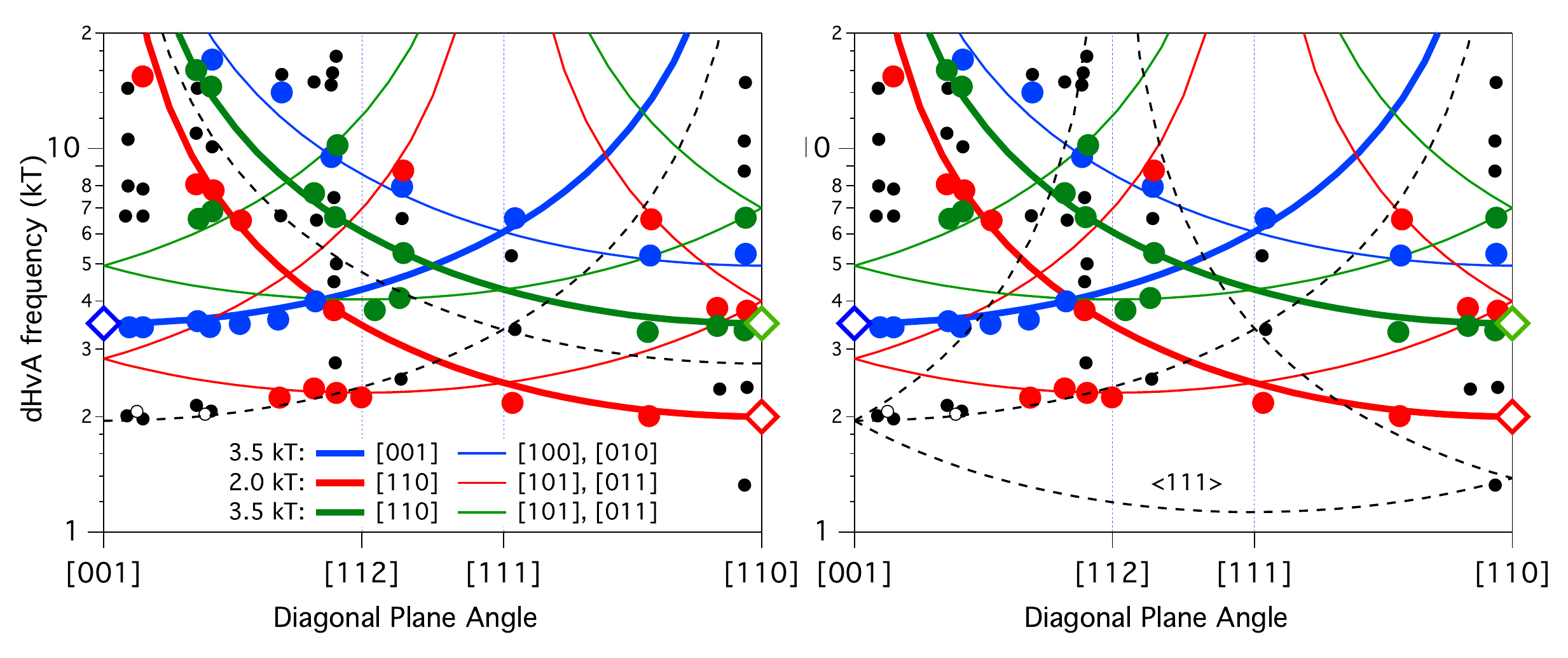}
\caption{
Possible additional 2D branches (dashed line) for the high frequency diagonal plane data.   (a)  $<$001$>$  surface state with 2 kT minimum contour.  (b) $<$111$>$ facet surface state with divergence along 35.3\deg\ but also a $<$2 kT minimum frequency branch.  
}
\label{extra10kT}
\end{figure*}

The ARPES-based 2D model of the high frequency diagonal plane in Fig. 4 of the main text successfully predicts more than 60\% of the angle-dependent dHvA frequencies.  Additional dHvA frequencies can  be understood by branches not currently observed in ARPES.  One possibility in Fig. \ref{extra10kT}(a) is an additional $<$001$>$ surface state with a 2.0 kT minimum frequency which can account for greater than 10 data points.   A second possibility in Fig. \ref{extra10kT}(b) to explain the [001] 2 kT frequencies, the $<$2 kT frequency along [110], and some $>$10 kT frequencies around 30\deg\ is the presence of $<$111$>$ facets with a $>$1 kT minimum frequency surface state.
